\documentclass[
 aip,
 amsmath,amssymb,
]{revtex4-1}

\usepackage{graphicx}
\usepackage{dcolumn}
\usepackage{bm}

\usepackage[utf8]{inputenc}
\usepackage[T1]{fontenc}
\usepackage{mathptmx}
\usepackage{etoolbox}

\makeatletter
\def\@email#1#2{%
 \endgroup
 \patchcmd{\titleblock@produce}
  {\frontmatter@RRAPformat}
  {\frontmatter@RRAPformat{\produce@RRAP{*#1\href{mailto:#2}{#2}}}\frontmatter@RRAPformat}
  {}{}
}%
\makeatother
\begin{document}


\title{Cavity correlations and the onset of charge ordering at charged interfaces: a modified Poisson-Fermi approach }
\author{Otávio David Braga}
 \affiliation{Instituto de F\'isica, Universidade Federal de Ouro Preto, 35400-000, Ouro Preto, MG, Brazil}
\author{Thiago Colla}%
 \email{colla@ufop.edu.br}
\affiliation{Instituto de F\'isica, Universidade Federal de Ouro Preto, 35400-000, Ouro Preto, MG, Brazil}

\date{\today}

\begin{abstract}
Charge layering in the close vicinity of charged interfaces is a well known effect, extensively reported in both experiments and simulations of Room Temperature Ionic Liquids (RTIL) and concentrated aqueous-like electrolytes. The traditional Poisson-Fermi (PF) theory is able to successfully describe overcrowding effects, but fails to reproduce charge ordering even in strong coupling regimes. Simple models, yet capable to investigate the interplay between these important interfacial phenomena, are still lacking. In order to bridge this gap, we herein present a modified PF  (mPF) approach which is able to capture layering effects in strong coupling regimes typical of RTIL. The modification is based on the introduction of charge cavities around test-particles, which simply extend the exclusion volume effects to also incorporate the accompanying depletion of charges due to particle insertion. Addition of this simple ingredient is shown to reproduce overscreening and charge ordering, thereby extending the predictive power of the PF approach to strong coupling regimes. Using a linear response theory, we are able to study the emergence of charge ordering based on two characteristic lengths: a wavelength responsible by charge layering, along with a dumping length which screens charge oscillations. At large ionic strengths and strong couplings, the system undergoes a transition to undumped charge layering. The transition takes place when the poles of the Fourier components of the linear potential become real-valued. This criteria  allows one to identify the transition line across the parameter space, thus delimiting the region of stability against unscreneed charge ordering.
\end{abstract}

\maketitle

\section{Introduction}\label{Intro}

Over the last decades, solvent-free electrolytes such as molten salts and ionic liquids have attracted  growing attention among the physical-chemistry community~\cite{Wilk02,Mac06,Fedo14,MacFarlane,Gaga22}, mostly due to their potential application in designing highly efficient electrochemical devices with tailoring properties~\cite{Mac07,Gho20,Cara21,Huan22,Wang22,Roper22,Wei23}. The increasing replacement of standard solvent-based electrolytes by such versatile liquids has been boosted by the development of modern techniques which allow for their ease of synthesis and reduced production cost~\cite{Lev02,Dee03,Burre07,Bey17,Sand20,Phil21,Kout22}. While molten salts are generally regarded as simple electrolytes characterized by high melting temperatures~\cite{Fedo14,Dupo11,Kry22}, ionic liquids are typically composed of complex molecular structures featuring high asymmetry~\cite{Hayes15}, which precludes crystal formation over a large temperature range. Such stability against crystallization represents a major practical advantage over molten salts in applications demanding operation over wide ranges of temperatures. Moreover, ionic liquids can be nowadays synthesized in a number of different molecular architectures~\cite{Phil21,Kout21,Llaver22}. Such a molecular-level resolution allows for a convenient bottom-up control of mesoscopic properties by changing quantities such as molecular shape and size~\cite{San11,Shap11,Nieder13,Yum13,Silva20,Kout23}, degree of asymmetry between cations and anions~\cite{Zhao19,Clark20,Andrei20,Souza22}, molecular polarizability~\cite{Res06,Shap11,Golo19}, affinity to an embedded surface~\cite{Sutto07,Daso21,Wei22}, among others. Besides, a number of ionic liquids are compatible with one another, such that they can be mixed up in different ways to provide a large variety of electrolytes with adjustable properties~\cite{Nieder12,Gol22,Elst23}. Of particular interest in the manufacturing of electronic and electrochemical devices are the so-called Room-Temperature Ionic Liquids (RTIL), which remain in the liquid phase over a wide range temperatures, spanning room-like temperatures and below ~\cite{Marsh02,Marsh04,Fedo14}. Apart from the large stability against crystallization, other clear advantages of RTIL rely on their ability to operate across large voltage windows~\cite{Maan13,Huan19},  non-volatility and high electrical conductivity~\cite{Maan13,Huan19,Gaga22}. Many RTIL are also environmental friendly, being considered as ``green solvent'', in strong contrast to most typical aqueous-like electrolytes~\cite{Mall12,Jin22,Sas23}. Due to these properties, RTIL have been extensively used in combination with electrode materials of high porosity and large conductivity in the designing of supercapacitors -- storage devices featuring enhanced performance, and capable of operating with large energy densities~\cite{Miao21,Saha24}. Giving the increasingly global need for sustainable, efficient and low cost storage/delivery of energy from reliable sources, the relevance of RTIL and related materials in the current development of electronic devices can not be underestimated. 

Despite the recent progress in the designing of novel RTIL-based devices and nanamaterials, advances in their theoretical description are still lagging behind in many aspects~\cite{Kon23}. The reason for such slow theoretical progress is manifold. The variety in shapes and complexity of molecular structures requires the use of ab-initio calculations with atomic-level resolutions~\cite{Tur03,Phil19}. As a result, large scale simulations able to capture the system thermodynamics over a large range of parameters (typical of experimental conditions) become computationally expensive. An alternative is to rely on coarse-graining models, where the missing microscopic details are somehow incorporated into effective parameters. Even at such mesoscopic level, the description is rather challenging. Compared to aqueous solutions, RTIL and solvent-free electrolytes can be characterized as weakly polar liquids, bearing typically low permittivities~\cite{Wak05,Fedo14}. As a result, electrostatic contributions can strongly dominate over entropic ones.  This is in contrast to what happens in polar liquids at room temperature, where the electrostatic self-energy of free ions rivals  their thermal energy~\cite{Levin09}.  In that sense, ionic liquids are similar to aqueous electrolytes at low temperatures, in which case the static structure is strongly dominated by charge ordering typical of electrolytes in the so-called strong coupling regime~\cite{Mor00,Bak11,Sam24}. Such an enhancement of electrostatic effects over thermal motion leads to a number of technical difficulties. Electrostatic correlations become strong, leading to the breakdown of traditional mean-field theories designed to describe aqueous-like electrolytes, such as the Poisson-Boltzmann (PB) and Gouy-Chapman (CG) approaches~\cite{Fedo14,Lev02}. The ionic distribution at the close vicinity of a charged interface -- the so-called electric double layer (EDL) -- predicted by these models features a rather simple structure. It is composed of a thin layer of condensed counterions, followed by a diffuse layer that fully screens the surface charge.~\cite{Phil96,Hansen00}. This simple scenario is no longer valid at strong coupling regimes, where combined packing and charge correlation effects play a non-trivial role in shaping the structure of the EDL~\cite{Korny07,Baza11,Agra24}. While packing constrains have an overall effect of limiting the amount of condensed counterions at the compact, inner layer, charge correlations favor the emergence of charge ordering. The latter effect is characterized by layers of alternating charges that pack together close to the interface, typically extending over many particle diameters into the bulk electrolyte, where the surface charge is finally neutralized~\cite{Lin12,Souza20,Avni20}. Taking account of strong correlation effects that lead to such complex EDL structures is a rather difficult task, requiring the employment of theoretical tools that go far beyond the mean-field level, such as field-theoretical approaches~\cite{Netz01,Hat09,Blos22} and classical density-functional theory~\cite{Wu11,Colla16,Kira21,Kon23}. Even in those cases, incorporation of electrostatic correlations in a numerically tractable way usually requires relying on approximation schemes like the random-phase approximation and related linear approaches, whose degree of accuracy in  situations of strong coupling regimes is rather questionable.  As an alternative, computer simulations can be employed to describe the ELD properties in RTIL with a fairly good degree of molecular resolution~\cite{Bed19,Kon23,Zhang24}. However, many technical aspects related to RTIL -- such as its complex molecular structure, high packing fractions and the long-range nature of Coulombic interactions  -- lead to much slower dynamics when compared to simulations of aqueous-like electrolytes~\cite{Gir17,Gir18}. For a comprehensive survey on the various aspects and challenges in modeling RTIL and key theoretical advances, we refer the reader to the seminal review of Fedorov and Kornyshev, see Ref. [\onlinecite{Fedo14}].

As is the case for simple electrolytes, it is rather useful to rely on simple models, yet able to shed light on the main physical mechanisms dictating the EDL structure in the case of RTIL. In that sense, lattice models have been widely employed with a reasonable degree of success to describe many features of RTIL in a physically transparent way~\cite{Dem12,Gir17,Gir18,Down18,Levy19}. Such approaches are particularly convenient to deal with densely packed configurations typical of EDLs at high applied voltages, as well as to model situations of high packing fractions typical of RTIL. One such successful model is the Poisson-Fermi (PF) approach~\cite{Korny07,Mar09,Fedo14} and its extensions~\cite{Fedo08,Xie16}, in which ions are meant to occupy lattice sites constrained to a single particle occupation, thus reproducing exclusion volume effects in a simple fashion. Instead of a Boltzmann distribution typical of a lattice gas with unconstrained site occupations~\cite{Fry18}, this approach leads to Fermi-like distributions that prevent the building-up of a unbounded counterion accumulation at the compact layer due to size effects. At sufficiently high applied voltages, the double layer structure is dominated by a counterion ``saturation effect'' in which counterions start to fill in subsequent layers close to the interface, until its charge is fully neutralized~\cite{Gir17}. Despite being somewhat artificial, this effect -- also referred to as \textit{overcrowding} -- is able to capture most of the physics behind packing effects, specially regarding the shape of capacitance curves~\cite{Fedo14}. The PF theory is however unable to predict charge ordering at the EDL, an effect closely related to an overscreening of the surface charge by the first layers of counterions. Recently, B\"ultmann and H\"arte proposed, in the context of density functional theory, the presence of regions around test-particles in which the electrostatic potential is constant, thus reproducing the effects of a charge neutral zone surrounding the test particle~\cite{Bul22}. This is accomplished by smearing out the ionic density over a spherical shell. A similar idea has been implemented by de Souza and co-workers~\cite{Souza20}, where different weight functions are assigned to size and electrostatic density functionals. These models are able to incorporate charge layering effect at moderate ionic couplings in a simple way.  Motivated by these works, we here propose a simple modification on the PF approach -- hereafter referred to as modified Poisson-Fermi approach (mPF) -- in which the exclusion volume of an ion is accompanied by cavity correlations that prevent charge penetration from neighboring ions. As we shall see, this simple modification is able to incorporate a large degree of charge ordering into the PF approach, specially in regions of strong couplings. The consequences of charge ordering and its relation to capacitance curves are explored in some detail. At very large couplings, it is shown that the theory becomes unstable in the sense that the range of charge ordering becomes unbound. Using a linear theory, we proceed to investigate the regions where such unstable solutions take place.

The remaining of the paper is organized as follows. In Section \ref{theory}, a small summary of the PF approach is presented, followed by the introduction of the new model, alongside with a  discussion on its stability considering a linear regime. Results for structure, capacitance curves and stability are presented in Section \ref{results}. Concluding remarks and perspectives for new approaches are then presented in Section \ref{close}.
 
\section{Theoretical background}\label{theory}

Before introducing cavity charge correlations into the PF description, it is instructive to briefly review the underlying framework of the lattice-gas model applied to ionic systems. We start by partitioning the space into a regular macroscopic lattice, which is then further splitted into smaller sub-lattices of equal size. Each sub-lattice is considered to be big enough to contain a large number of sites $V$, but yet small in comparison to typical system dimensions. The position of each sub-lattice is represented by a vector $\bm{r}=a(n_x\bm{\hat{e}}_x+n_y\bm{\hat{e}}_y+n_z\bm{\hat{e}}_z)$, where $a$ is a typical dimension of a sub-lattice, and the $n_i$'s are integers. The configuration of a sub-lattice at $\bm{r}$  is fully described by specifying the number of cations $N_{+}(\bm{r})$ and anions $N_{-}(\bm{r})$ it contains. All sub-lattices are considered to be in chemical equilibrium, which is achieved once each cell contains the proper number of particles to ensure a constant chemical potential throughout the system. Our goal is to determine the free-energy density associated with each sub-lattice, from which macroscopic thermodynamic information can be extracted by invoking a local approximation. To this end, let us focus on a given sub-lattice containing $N_{\pm}$ cations/anions. The site dimensions are chosen such that each site can accommodate one particle at most. This constrain is aimed to incorporate volume exclusion effects, as particles are not allowed to overlap each other.  The associated number of microscopic configurations can be easily obtained considering  a three-state model which assigns each lattice site a tag ``$+$'', ``$-$'' or ``$0$'', depending on whether the site is occupied by a cation, an anion or empty, respectively~\footnote{We recall that an empty site can be regarded as representing either an empty space (e. g., a void) or a solvent component.}. The number of distinct configurations is clearly given by the counting factor
\begin{equation}
    \Omega(\{N_i\})=\dfrac{V!}{N_{+}!N_{-}!(V-N_{+}-N_{-})!}.
    \label{Omega}
\end{equation}
The corresponding entropy contribution per unit of site $s\equiv S/V =k_B\ln(\Omega)/V$ (where $k_B$ is the Boltzmann constant) to the lattice-gas can be easily obtained by invoking Stirling approximation in the limit $V\gg 1$:
\begin{equation}
   s=k_B\left[\phi_{+}\ln(\phi_{+})+\phi_{-}\ln(\phi_{-})+(1-\phi_{+}-\phi_{-})\ln\left(1-\phi_{+}-\phi_{-}\right)\right],
   \label{S1}
\end{equation} 
where $\phi_{\pm}=N_{\pm}/V$ is the volume fraction of cations and ions assigned to the sub-lattice. The energy associated with these configurations can be estimated in the context of a mean-field approach, in which an average electrostatic potential $\psi_{\pm}$ (yet to be determined) acts upon cations and anions belonging to the sub-lattice. The corresponding energy is then $U=(N_{+}\psi_{+}-N_{-}\psi_{-})$, such that the energy density on the sub-lattice reads
\begin{equation}
   u=U/V=\phi_{+}\psi_{+}+\phi_{-}\psi_{-}.
   \label{u}
\end{equation}
In this mean-field approach, the potentials $\psi_{\pm}$ are considered as the averaged quantities resulting from charges located inside and outside the sub-lattice (including external sources) when a cation/anion is placed at the sub-lattice. The free-energy density $f=F/V=u-Ts$ can now be readily computed from Eqs. (\ref{S1}) and (\ref{u}) as a function of local packing fractions:
\begin{equation}
   \beta f(\phi_{+},\phi_{-})=\phi_{+}\beta\psi_{+}+\phi_{-}\beta \psi_{-}+\phi_{+}\ln(\phi_{+})+\phi_{-}\ln(\phi_{-})+(1-\phi_{+}-\phi_{-})\ln\left(1-\phi_{+}-\phi_{-}\right),
   \label{f}
\end{equation}
where $\beta=1/k_BT$. At this point, a coarse-graining description can be employed in which the overall free-energy is computed as a sum of $f(\phi_{+}(\bm{r}),\phi_{-}(\bm{r}))V$ evaluated at each sub-lattice. In the limit when $V$ is considered to be much smaller than the total number of sites, a continuum approximation can be invoked. The resulting free-energy will be a functional of the local densities, and the equilibrium distributions can be calculated in the context of a classical density-functional theory, once the system is properly connected to a reservoir of fixed chemical potentials~\cite{hansen90a}. Alternatively, the local packing fractions $\phi_{\pm}$ can be taken as variational parameters that minimize the free-energy densities on each sub-lattice, subject to the constrain of chemical equilibrium between the various sub-systems. The physics behind this assumption is that particles can be freely exchanged between different sub-lattices, until an overall chemical equilibrium of vanishing particle flux is achieved. This requirement results in the following Euler-Lagrange conditions for the equilibrium packing fractions
\begin{equation}
  \dfrac{\partial f(\phi_{+},\phi_{-})}{\partial\phi_{\pm}}=\mu,
  \label{EL}
\end{equation}
where $\mu$ is the chemical potential that guarantees chemical equilibrium between the sub-systems, considering symmetry between positive and negative species (which is the case either for an electrolyte confined between appositely charged walls of same magnitude, or in contact with a reservoir of fixed concentration, to be considered later on). Application of the condition (\ref{EL}) in Eq. (\ref{f}) leads to following pair of algebraic equations
\begin{eqnarray}
    \dfrac{\phi_+}{1-\phi_+-\phi_-} & = & e^{\beta(\mu-q\psi_+)},\label{phi01}\\
    \dfrac{\phi_-}{1-\phi_+-\phi_-} & = & e^{\beta(\mu+q\psi_-)}.\label{phi02}  
\end{eqnarray}
This set of coupled equations can be easily solved, resulting in the following equilibrium distributions
\begin{equation}
  \phi_{\pm}(\bm{r})=\dfrac{\eta e^{\mp\beta q\psi_{\pm}(\bm{r})}}{1+\eta\left(e^{-\beta q\psi_{+}(\bm{r})}+e^{\beta q\psi_{-}(\bm{r})}\right)},
  \label{phi1}
\end{equation}
where $\eta\equiv e^{\beta\mu}$ and $q$ is the elementary charge. These Fermi-like distributions satisfy the condition $0<\phi_{\pm}\le 1$, which reflects the single-site occupation condition, in much the same way as the occupation pattern of energy levels in a Fermi gas.  Notice that the local character of both packing fraction and averaged potentials has been emphasized by their dependency on the sub-lattice coordinate $\bm{r}$. Now, a continuum limit can be employed in which the sub-lattice dimension $a$ is much smaller than the typical system size, and the corresponding volume $V\sim a^3$ becomes negligible in comparison to the system volume. In this case, the position $\bm{r}$ of a sub-lattice can be taken as a continuum parameter in Eq. (\ref{phi1}). This approximation is a rather convenient one, since it allows one to easily relate the macroscopic potentials to the average local concentrations.  Moreover, taking into account that the system is in the presence of flat, uniformly charged surfaces along the $xy$ plane, translation symmetry along this plane allows one to further identify $\bm{r}\rightarrow z$. If the system is in contact with a reservoir of bulk packing fraction $\phi_b$, the condition $\phi_{\pm}(z\rightarrow\infty)=\phi_b$ implies  $\eta=2\phi_b/(1+\phi_b)$ and thus
\begin{equation}
  \phi_{\pm}({z})=\dfrac{\phi_b e^{\mp\beta q\psi_{\pm}({z})}}{1+\phi_b\left(e^{-\beta q\psi_{+}({z})}+e^{\beta q\psi_{-}({z})}-2\right)}.
  \label{phi2}
\end{equation}
This is the situation of a single-wall system. In the case where system is confined between two oppositely charged walls of surface charge density $\sigma$, it is sometimes more convenient to fix the overall electrolyte concentrations instead.

In the original formulation of the mean-field lattice gas model~\cite{Korny07}, no distinction is made between potentials around cations and anions. In this case, the quantities $\psi_{\pm}$ are identified with the average potential $\psi(z)$, and the distributions (\ref{phi2}) become

\begin{equation}
  \phi_{\pm}({z})=\dfrac{\phi_b e^{\mp\beta q\psi({z})}}{1+\phi_b\left(\cosh(\beta q\psi)-1\right)}=\dfrac{\phi_b e^{\mp\beta q\psi({z})}}{1+\phi_b\sinh^2\left(\dfrac{\beta q\psi}{2}\right)}.
  \label{phi3}
\end{equation}
The average potential is calculated self-consistently by means of the Poisson equation,
\begin{equation}
 \dfrac{d^2\psi(z)}{dz^2}=-\dfrac{4\pi q}{\varepsilon}\left[v\phi_{+}(z)-v\phi_{-}(z)+\varrho_w(z)\right],
  \label{poisson1}
\end{equation}
where $v$ is the volume of an ion, $\varepsilon$ is the electrolyte permittivity, and $\varrho_w(z)$ represent the charge-density assigned to the fixed walls. 

In the close vicinity of a strongly charged interface, the potential $\psi$ becomes very large in magnitude, and the packing fractions of cations (anions) in (\ref{phi3}) approaches unity in the case of negatively (positively) charged surfaces. As a result, the double layer becomes ``saturated'', as all lattice sites adjacent to the wall become occupied by neutralizing counterions. This well-known result strongly contrasts the unbounded growth in counterion concentration -- way beyond the close-packing condition -- predicted by point-like ion approaches (\textit{e. g.}, Poisson-Boltzmann theory). This qualitative change in double layer structure has a dramatic impact in the shape of underlying capacitance curves~\cite{Fedo14}. While point-like ion models predict a ``U'' shape capacitance with a unphysical increase of capacitance at high potentials, the PF model predicts either ``camel-like'' or ``bell shaped'' curves that correctly account for the capacitance decrease at strongly charged surfaces~\cite{Fedo14}. Despite this clear improvement, the counterion saturation at high applied voltages seems to be unrealistic in view of electrostatic correlations that should prevent the formation of long-range structures of same net charge. In what follows, we shall propose a simple modification of the PF approach which allows to include charge correlations in the same way as size effects are taken into account, namely by the introduction of charge cavities that prevent particle overlap. 

\subsection{Charge cavity correlations}

In order to motivate our model of charge cavity correlations, we start with some general considerations regarding the effects of creating a cavity in an otherwise uniform system of particles and charges, in the context of a mean-field approach. To this end, we consider a particle insertion method whose arguments go along the lines of the classical scaled particle theory~\cite{Reiss59,Gibb69}. First, let us consider the simplest case of inserting a \textit{point} particle in a system of \textit{uncharged} particles. The cost of such a particle insertion entails only an entropic penalty proportional to the particle concentration of the specie under consideration, \textit{i. e.}, an ideal gas contribution $\beta\mu_i=\ln(\phi_i)$. The requirement of a fixed overall chemical potential $\mu$ throughout the system should then result in a uniform concentration, say $\bar{\phi}_i$. There will be a penalty upon particle insertion when $\phi_i>\bar{\phi}_i$, whereas in regions where $\phi_i<\bar{\phi}_i$ particle insertion will be favorable. Apart from such entropic contribution, the gain/penalty resulting from a point-particle insertion depends on the change of energy due to the interaction with other species~\cite{hansen90a}. These are the basic ingredients that dictates equilibrium condition in a mean-field theory of point-like particles. 

Now, let us consider that particle insertion is also accompanied by the formation of a cavity, whose size scales with particle dimensions. Insertion of a hole requires an energy cost $\Delta U_0$, due to exclusion volume effects from particles already present into the system. The probability $P_0$ of particle insertion is related to the energy change through the Boltzmann factor $P_0=e^{-\beta\Delta U_0}$. In a mean-field level, the probability of a successful particle insertion can be taken as the volume fraction of empty space, $\phi_0=1-\sum_i\phi_i$, where the sum is computed over all species. The related energy change upon particle insertion is thus
\begin{equation}
    \beta\Delta U_0=\ln\left(\dfrac{1}{1-\sum_i\phi_i}\right).
    \label{mui1}
\end{equation}

The free-energy change is now comprised of entropic and exclusion volume contributions $\beta\mu_i=\ln(\phi_i)+\beta\Delta U_0$, and thus reads 
\begin{equation}
    \beta\mu_i=\ln\left(\dfrac{\phi_i}{1-\sum_j\phi_j}\right).
    \label{mui1}
\end{equation}
Clearly, this contribution is closely related to the terms on the left-hand side of Eqs. (\ref{phi01}) and (\ref{phi02}). These are the basic mechanisms behind exclusion volume effects at the PF approach. In the case of an overcrowded environment, the contribution $1-\sum_j\phi_j$ in (\ref{mui1}) approaches zero, and the cost for particle addition becomes extremely large. As a result, in positions where particle insertion becomes energetically favorable (\textit{e. g.}, counterions next to a charged wall), the contribution above will eventually prevent the building-up of particle concentrations exceeding close-packing. 

Let us now focus on the change of electrostatic energy corresponding to a particle insertion, with particular emphasis on the role played by cavity formation. For simplicity, we consider a spherical region of radius $a$ (as a sub-lattice defined in the previous section). The charges therein are smeared out into  a rigid background of uniform charge density $\rho_b$. The electrostatic self-energy of such a configuration can be readily evaluated to be $U_{self}=3Q_b^2/5\varepsilon a$. If a point charge of magnitude $q$ is now inserted at the center of the sphere, the electrostatic energy is changed due to its interaction with the rigid background charge\footnote{If the background charge does not contain external sources, fluctuations resulting from a charge insertion would lead to a polarization of the background charge. Since our aim is only to investigate the effects of a cavity insertion, we consider that the background charge is unaffected by the insertion of an extra charge.}. The corresponding change in electrostatic energy due to the interaction between background and point-charge is
\begin{equation}
    \Delta U_{elec}=\int\left(\dfrac{q\delta(r)}{4\pi r^2}\right)\psi_b(r)d\bm{r}=q\psi_b(0)=\dfrac{3Q_bq}{2\varepsilon a},
\label{Del_U1}
\end{equation}
where $Q_b=\dfrac{4\pi a^3}{3}\rho_b$ is the total background charge, and $\psi_b(r)=\dfrac{3Q_b}{2\varepsilon a}\left(1-\dfrac{r^2}{3a^2}\right)$ the corresponding potential. If test and background charges have same sign, particle insertion clearly results in an energy penalty, while insertion of an opposite charge becomes energetically favorable. This manifests the trend of a charged system to acquire local electroneutrality. Equilibrium is then dictated by a local balance between entropy -- which attempts to avoid inhomogeneous distributions --  and the electrostatic tendency to establish charge neutrality. It is important to point out that the mean-field lattice model outlined in the previous section properly accounts for exclusion effects at the entropic level, but the underlying electrostatics is that of point ions interacting with background charges. In order to evaluate the effects of cavity formation at the electrostatic interactions, let us now suppose that the charge $q$ is centered at a spherical cavity of radius $d$, which prevents the penetration of external charges. The background charge is now constrained to occupy the region $d<r<a$. The potential due to the background at the particle position is now reduced to $\psi_b=\dfrac{2\pi\rho_b a^2}{\varepsilon}\left(1-\dfrac{d^2}{a^2}\right)$, such that the interaction energy between the added charge and the fixed background now reads
\begin{equation}
    \Delta U'_{elec}=q\psi_b(0)=\dfrac{3Q_{b}q}{2\varepsilon a}\left(1-\gamma^2\right),
\label{Del_U2}
\end{equation}
where $\gamma\equiv d/a$. Creation of a cavity thus reduces the magnitude of energy change by a factor  $(d/a)^2$. Likewise, a straightforward calculation shows that the self-energy of the rigid background becomes $U_{self}'=\dfrac{3Q_b}{5\varepsilon a}(1-5\gamma^3/2+3\gamma^5/2)$. Since $\gamma<1$ the dominant contribution comes from particle/background interaction, Eq. (\ref{Del_U2}). Addition of a hole thus reduces (increases) the energy gain (penalty) upon insertion of a particle of opposite (same) sign into the background. In other words, the ability to neutralize the effect of fixed charges becomes less effective. The net result is that a larger amount of mobile charge becomes necessary to compensate the electrostatic energy of an external source of opposite charge. As we shall see, these effects from a charge cavity insertion will have a strong impact in double layer structure around a charged interface. 

Having established the qualitative effects of a charge cavity, we are now in position to incorporate these effects into the mean-field PF approach. The basic idea is to consider that the averaged potentials $\psi_{\pm}(\bm{r})$ in Eqs. (\ref{phi1}) and (\ref{phi2}), resulting from insertion of a catio/anion at $\bm{r}$, comprise the contribution from a charge cavity centered at this position. The potentials then read
\begin{equation}
\psi_{i}(\bm{r})=\sum_{j=\pm}\dfrac{q}{\varepsilon}\int \dfrac{\alpha_j\rho_j(\bm{r}')\Theta(|\bm{r}-\bm{r}'|-d_{ij})}{|\bm{r}-\bm{r}'|}d\bm{r}'+\int\dfrac{\varrho_w(\bm{r}')}{\varepsilon|\bm{r}-\bm{r}'|}d\bm{r}',
\label{psi1}
\end{equation}
where $\Theta(x)$ represents the unit-step function, which equals one if $x>0$ and zero otherwise. The second term above describes the potential due to the fixed walls. The step function alongside the ionic profiles in the first integral guarantees that the electrostatic potential felt by an ion $i$ at $\bm{r}$ is produced only by source charges $j$  at points $\bm{r}'$ lying beyond the cavity of size $d_{ij}$ around the observation point. It is convenient to split these potentials into mean-field and fluctuation contributions, $\psi_{\pm}(\bm{r})=\psi^{mf}(\bm{r})+\psi^{fluc}_{\pm}(\bm{r})$. The mean-field potential is
\begin{equation}
\psi^{mf}(\bm{r})=\sum_{j=\pm}\dfrac{q}{\varepsilon}\int \dfrac{\alpha_j\rho_j(\bm{r}')}{|\bm{r}-\bm{r}'|}d\bm{r}'+\int\dfrac{\varrho(\bm{r}')}{\varepsilon|\bm{r}-\bm{r}'|}d\bm{r}',
\label{psi_mf1}
\end{equation}
and clearly corresponds to the point-like ion contribution typical of mean-field models. The fluctuation potential encompasses the contribution from a cavity formation, and is given by
\begin{equation}
 \psi_{i}^{fluc}(\bm{r})=-\sum_{j=\pm}\int \alpha_j\dfrac{\rho_j(\bm{r}')\Theta(d_{ij}-|\bm{r}-\bm{r}'|)}{|\bm{r}-\bm{r}'|}d\bm{r}'. \label{psi_f1}
\end{equation}
The fluctuation potential is related to charge fluctuations upon insertion of a particle at a test point. Therefore, the cavity correlation $\Theta(d_{ij}-|\bm{r}-\bm{r}'|)$ should be in practice recognized as the total correlation function $h_{ij}(\bm{r},\bm{r}')$ between ionic species~\cite{Levin02,hansen90a}. Accurate calculation of this quantity in the case of inhomogeneuos systems is a very complex task, which often requires access to three-body correlations or beyond. In this case, the present approach can be interpreted as an approximation which only retains an exclusion cavity contribution to particle correlations. The size of this region can generally depend on local concentrations and the exclusion volume upon close contact. For example, the exclusion zone separating cations and anion should be proportional to their closest-contact distance, whereas cation-cation or anion-anion exclusion can be considered larger due to their mutual repulsion. 

In the case of electrolytes in contact with flat plates bearing uniform surface charges and lying along the $xy$ plane, the in-plane integrals can be easily computed. As a result, the mean-field potential, Eq. (\ref{psi_mf1}) is simplified to
\begin{equation}
    \psi^{mf}(z)=\dfrac{2\pi q\sigma}{\varepsilon}(L-2z)+\dfrac{2\pi q}{\varepsilon}\sum_{j=\pm}\alpha_j\left[z\left(2I_{0j}(z)-I_{0j}(L)\right)-2I_{1j}(z)+I_{1j}(L)\right],
    \label{psi_mf2}
\end{equation}
where $\sigma$ is the surface charge density, and $L$ is the distance between the oppositely charged plates. The functions $I_{0j}(z)$ and $I_{1j}(z)$ are defined as
\begin{eqnarray}
I_{0j}(z) & = & \dfrac{1}{v}\int_0^{z}\phi_j(z')dz',\label{I0}\\
I_{1j}(z) & = & \dfrac{1}{v}\int_0^{z}z'\phi_j(z')dz',\label{I1}
\end{eqnarray}
and are related to the net and dipole charge contributions, respectively, to the electrostatic potential. In terms of these quantities, the fluctuation potential can be written as
\begin{equation}
\begin{split}
    \psi^{fluc}_i(z)=\dfrac{2\pi q}{\varepsilon}\sum_{j=\pm}\alpha_j\biggr[z\left(2I_{0j}(z)-I_{0j}(z-d_{ij})-I_{0j}(z+d_{ij})\right)-d_{ij}\left(I_{0j}(z+d_{ij})-I_{1j}(z-d_{ij})\right)\\
    +\left(I_{1j}(z+d_{ij})+I_{1j}(z-d_{ij})-2I_{1j}(z)\right)\biggr].
    \end{split}
    \label{psi_f2}
\end{equation}
Together, Eqs. (\ref{phi2}), (\ref{psi_mf2}) and (\ref{psi_f2}) provide a closed set of integral equations that allows for a simultaneous calculation of ionic profiles and averaged potentials. The numerical solution can be achieved by means of a simple Picard iteration as follows. Starting with guess functions for the local packing fractions $\phi_{\pm}(z)$, Eqs. (\ref{I0}) and (\ref{I1}) are numerically evaluated, and the average potential is then computed from Eqs. (\ref{psi_mf2}) and (\ref{psi_f2}). New profiles are then obtained from Eq. (\ref{phi2}), and the whole process is repeated until convergence is achieved. 

\subsection{Linear regime}\label{linear}

Many features of the proposed mPF approach can be better understood by investigating its linear regime. Far away from the charged plates, the average potential becomes rather small ($\beta q \psi\ll 1$), and Eqs. (\ref{phi2}) can be expanded up to first order in the potentials $\psi_{\pm}(z)$. The local packing fractions in this limit thus become
\begin{equation}
    \phi_{i}(z)\approx\phi_b[1-\beta q\psi_{i}(z)].
    \label{phi_lin}
\end{equation}
To work out this linear regime, it is convenient to write down the potentials in reciprocal space. Defining the Fourier components of $\psi_{i}(\bm{r})$ as 
\begin{equation}
\hat{\psi}_{i}(\bm{k})=\int \psi_{i}(\bm{r})e^{i\bm{k}\cdot\bm{r}}d\bm{r},
\label{psi_k1}
\end{equation}
equation (\ref{psi1}) can be simplified to
\begin{equation}
    \hat{\psi}_{i}(\bm{k})=\dfrac{qv}{\varepsilon}\sum_{j}z_j\hat{\phi}_j(\bm{k})G_{ij}(\bm{k})+\dfrac{4\pi\hat{\varrho}_w(\bm{k})}{\varepsilon k^2},
    \label{psi_k2}
\end{equation}
where the coefficients $G_{ij}(\bm{k})$ are defined as
\begin{equation}
    G_{ij}(\bm{k})=\int \dfrac{\Theta(r-d_{ij})}{r}e^{i\bm{k}\cdot\bm{r}}d\bm{r}.                           \label{Gij1}
\end{equation}
This integral can be easily evaluated, yielding
\begin{equation}
    G_{ij}(k)=\dfrac{4\pi}{k^2}\cos(kd_{ij}). 
    \label{Gij2}
\end{equation}
These coefficients represent an  electrostatic interaction between ions of types $i$ and $j$ that is ``turned on'' only when their mutual separation is larger than the cavity size $d_{ij}$.  Since both potentials and charge densities depend only on the transversal coordinate $z$, the Fourier components in (\ref{psi_k1}) are simplified to 
\begin{equation}
\hat{\psi}_{i}(\bm{k})=(2\pi)^2\delta^2(\bm{k}_s)\int_{0}^{L}\psi_i(z)e^{ik_zz}dz,
\label{psi_k3}
\end{equation}
where $\bm{k}_s\equiv k_x\bm{\hat{e}}_x+k_y\bm{\hat{e}}_y$ is the in-plane wavevector. In particular, the Fourier components $\hat{\varrho}_w(\bm{k})$ of the wall charge density $\varrho_w(z)=\sigma q[\delta(z)-\delta(z-L)]$ are
\begin{equation}
\hat{\varrho}_w(\bm{k})=\delta(\bm{k}_s)\sigma q\left(1-e^{-ik_zL}\right).
\label{rhow_k}
\end{equation}
As a result, Eq. (\ref{psi_k1}) depends non-trivially only on the transverse wavevector $k_z$, which can be further identified with $k_z=|\bm{k}|=k$. Using these results, Eq. (\ref{psi_k2}) can be re-written more explicitly as
\begin{equation}
   \hat{\psi}_i(k)=\dfrac{4\pi q}{\varepsilon k^2}\left[\sigma(1-e^{ikL})+v\sum_jz_j\hat{\phi}_j(k)\cos(kd_{ij})\right].
   \label{psi_k4}
\end{equation}
So far, this result is exact, and simply expresses the coulomb potential with a cavity correction (the Poisson equation for the two-plate system is readily recovered in the limit $d_{ij}\rightarrow 0$). Now the linear approximation can be invoked by considering the Fourier transform of Eq. (\ref{phi_lin}), $\hat{\phi}_i(k)=\phi_b[\delta(k)-\beta q\hat{\psi}_i(k)]$. Substitution into Eq. (\ref{psi_k4}) leads to the following set of algebraic equations in the linear regime:
\begin{equation}
   \hat{\psi}_i(k)=\dfrac{4\pi q}{\varepsilon k^2}\left[\sigma(1-e^{ikL})-\beta qv\phi_b\sum_jz_j^2\hat{\psi}_j(k)\cos(kd_{ij})\right],
   \label{psi_k5}
\end{equation}
where usage has been made of the electroneutrality condition $\sum_i \phi_bz_i=0$ in the limit $k\rightarrow 0$. This set of equations can be solved to obtain the linear potentials $\hat{\psi}_{\pm}(k)$. Since our main concern is to analyse the onset of charge ordering, let us consider the case where all cavities are equally sized, $d_{ij}=d$, which significantly simplifies the algebraic steps towards the solution of Eq. (\ref{psi_k5}). Also taking into account that $z_i=\pm$, it follows that $\hat{\psi}_{+}(k)=\hat{\psi}_{-}(k)$, and the coefficients have to satisfy
\begin{equation}
     \hat{\psi}_{\pm}(k)=\dfrac{4\pi q }{ \varepsilon k^2}(1-e^{ikL})-\dfrac{\kappa^2\cos(kd)}{k^2}\hat{\psi}_{\pm}(k) ,
   \label{psi_k6}  
\end{equation}
where $\kappa\equiv\sqrt{8\pi\lambda_B\phi_bv}$ defines the inverse Debye length at the bulk electrolyte, and $\lambda_B=\beta q^2/\varepsilon$ denotes the Bjerrum length. The linear potentials in the case of monovalent ions with equally sized electrostatic cavities thus become
\begin{equation}
     \beta q\hat{\psi}_{\pm}(k)=\dfrac{4\pi\lambda_B (1-e^{ikL})}{ k^2+\kappa^2\cos(kd)}.
   \label{psi_k7}  
\end{equation}
In real space, the potential is calculated via the inverse transform of Eq. (\ref{psi_k3}), namely
\begin{equation}
     \hat{\psi}_{i}(z)=\dfrac{1}{2\pi}\int_{-\infty}^{\infty}e^{-ikz}\hat{\psi}_{i}(k)dk.
   \label{psi_z1}  
\end{equation}
Substitution of (\ref{psi_k7}) into (\ref{psi_z1}) shows that the linear potential can be decomposed as $\psi(z)=\varphi(z)-\varphi(z-L)$, where the function $\varphi(z)$ is given by
\begin{equation}
    \varphi(z)=2\lambda_B\int_{-\infty}^{\infty}\dfrac{e^{-ikz}}{k^2+\kappa^2\cos(kd)}dk.
    \label{phi_z2}
\end{equation}
A formal solution of this integral can be achieved by an analytical continuation in which the $k$ values are extended over the upper complex plane.  Straightforward application of the Residue Theorem~\cite{churchill09} allows one to write the formal solution as
\begin{equation}
    \varphi(z)=4\pi i\lambda_B \sum_j Res(k_j),
    \label{phi_z3}
\end{equation}
 where the sum is performed over the residues of (\ref{phi_z2}) across the upper half-plane. Since the function $e^{-ikz}$ is analytic throughout the plane, the poles are determined by the zeros of $k^2+\kappa^2\cos(kd)$ along the upper plane. If $k_{j}=k_{Rj}+ik_{Ij}$ denotes the $j$th root, its real $k_{Rj}$ and imaginary $k_{Ij}$ parts are bound to satisfy the set of algebraic equations
 \begin{eqnarray}
     k_{Rj}^2 - k_{Ij}^2+\kappa^2\cosh(k_{Ij}d)\cos(k_{Rj}d) & = & 0\label{kj_1},\\
     2k_{Rj}k_{Ij}-\sinh(k_{Ij}d)\sin(k_{Rj}d) & = & 0.\label{kj_2}
 \end{eqnarray}
It is clear from these equations that the roots $k_j$ are symmetric with respect to the imaginary axis. Furthermore, the potential $\phi(z)$ in (\ref{phi_z2}) is clearly real-valued, as the imaginary part is evaluated as an symmetric integral of an odd function. In the absence of real roots, Eq. (\ref{phi_z3}) can thus be written quite generally as 
 \begin{equation}
    \varphi(z)=4\pi \lambda_B \sum_j A_j(z)e^{-k_{Ij}z}\cos(k_{Rj}z+\alpha_j),
    \label{phi_z4}
\end{equation}
where $A_j(z)$ is a polynomial in $z$ which depends on the order of the pole~\cite{churchill09}, and $\alpha_j$ is a phase shift. The summation is performed over the roots lying along the first quadrant of the complex plane. This equation, in combination with Eq. (\ref{phi_lin}), clearly demonstrates the onset of charge ordering, whose decay and oscillating modes are determined from solutions of  Eqs. (\ref{kj_1}) and (\ref{kj_1}). In the asymptotic limit $z\rightarrow \infty$ the dumping oscillations are related to the roots of (\ref{kj_1}) and (\ref{kj_2}) which lie closest to the real axis. In particular, in the limit $d\rightarrow 0$, the only root becomes $k=i\kappa$ and Eq. (\ref{phi_z4}) recovers the Debye-H\"uckel potential $\phi(z)=4\pi\lambda_B\sigma e^{-\kappa z}$, with a monotonic decay dictated by the Debye screening length $\zeta=\kappa^{-1}$. Clearly, the presence of an electrostatic cavity induces harmonic oscillations typical of charge layering, whose decay depend not only on the screening parameter $\kappa$, but also on the cavity size $d$ in a non-trivial way.

One possible root of Eqs. (\ref{kj_1}) and (\ref{kj_2}) corresponds to $k_{Ij}=0$, in which case the root of $k^2+\kappa^2\cos(kd)$ crosses the real axis. In this case, the path along the real axis must be either shifted or deformed to avoid the singular point~\cite{churchill09}, and Eq. (\ref{phi_z3}) will contain an extra contribution from the contour change. This contribution can be evaluated by defining a semi-circle of radius $\epsilon$ which circumvents the real pole from above, and then taking the limit $\epsilon\rightarrow 0$. Let $k_0$ be the real root of Eq. (\ref{kj_1}) with $k_{Ij}=0$, then the contour can be parameterized as $k\rightarrow k_0+\epsilon e^{i\theta}$, with $0\le \theta\le \pi$, and the factor $k^2+\kappa^2\cos(kd)$ can be approximated along the path as $(2k_0-\kappa^2\sin(k_0d))\epsilon e^{i\theta}$. The contour integral $I(k_0)$ thus becomes
\begin{equation}
    I(k_0)=ie^{-ik_0z}\epsilon \int_0^{\pi}\dfrac{e^{-i\epsilon\cos\theta z}e^{-\epsilon\sin\theta z}e^{i\theta}}{\epsilon  e^{i\theta}(2k_0-\kappa^2\sin(k_0d))}d\theta
    \label{Ie1}
\end{equation}
which, upon taking the limit $\epsilon\rightarrow 0$, reduces to
\begin{equation}
    I(k_0)=\dfrac{ie^{-ik_0z}}{2k_0-\kappa^2\sin(k_0d)}\int_0^\pi d\theta=\dfrac{i\pi e^{-ik_0z}}{2k_0-\kappa^2\sin(k_0d)}.
    \label{Ie1}
\end{equation}
This contribution can be combined with the contour integral $I(-k_0)$ surrounding the accompanying negative root $-k_0$, thus adding the following contribution to the integral over the real axis:
\begin{equation}
    I(k_0)+I(-k_0)=\dfrac{2\pi \sin(k_0z)}{\kappa^2\sin(k_0d)-2k_0}.\label{Ie2}
\end{equation}
If there are $n$ such real roots, each will add a contribution like (\ref{Ie2}) to the integral in (\ref{phi_z3}), in addition to the damping decay of (\ref{phi_z4}). Note that the potential does not decay exponentially, even in the far-field limit. Instead, the charge layering is extended over the entire system. Application of the Residue Theorem now leads to the general result
\begin{equation}
    \varphi(z)=4\pi \lambda_B\sigma \left[\sum_{\alpha}A_{0\alpha}\sin(k_{0\alpha}z)+\sum_j A_j(z)e^{-k_{Ij}z}\cos(k_{Rj}z+\alpha_j)\right],
    \label{phi_z5}
\end{equation}
where the first sum is extended over the real roots. The appearance of unscreened oscillations can have a physical realization only in the case of a canonical formalism in which the overall particle number confined by two plates is known. It becomes meaningless in the case of open systems, where the far-distance profiles have to relax to their bulk values.  

The region in parameter space where such unstable open-system solutions start to emerge can be identified by investigating the real roots of the algebraic equation
\begin{equation}
\tilde{k}_0^2+\tilde{\kappa}^2\cos(\tilde{k}_0)=0,
\label{roots}
\end{equation}
where $\tilde{k}_0\equiv k_0d$ and $\tilde{\kappa}\equiv \kappa d$.  In what fallows, we shall use this criteria to investigate the region of stable solution over the parameter space.  

\section{Results}\label{results}

We now apply the model presented above to study the structure of the double layer around a flat, charged interface, comparing its predictions with the traditional PF approach. We consider a symmetric electrolyte with equally sized, monovalent cations and anions of radius $r_{ion}=0.25$~nm, at room temperature.
In what follows, we shall consider two representative electrolytes with dielectric constants $\varepsilon_r=80$, typical of an aqueous electrolyte, and $\varepsilon_r=18$, lying in a range more characteristic of ionic liquids. A measure of the strength of electrostatic correlations in such systems is provided by the coupling parameters $\Gamma_{ij}=\beta \alpha_i\alpha_jq^2/\varepsilon r_{ion}=\alpha^2\lambda_B/2r_{ion}$. These parameters quantify the ratio between electrostatic and thermal energies at closest ionic contact. At low temperatures and/or high permittivity, electrostatic interactions prevail over thermal motion, and the system structure becomes dominated by charge ordering, in opposition to particle-particle correlations (such as packing effects). In the present situation of a symmetric electrolyte, the coupling parameter becomes $\Gamma=2.88$ and  $\Gamma=13.6$ for the aqueous and ionic liquid based electrolytes, respectively, allowing us to investigate the role played by charge ordering in the double layer structure and underlying capacitance curves. We consider the plate separation $L$ to be large enough so that a well defined bulk packing fraction $\phi_b$ is established far away from the plates. Below, we compare PF and mPF predictions for the EDL structure and capacitance curves. Moreover, we analyse in detail the emergence of undumped oscillations in the mPF model. 

\subsection{Structure of the EDL}

\begin{figure}[h!]
    \centering
    \includegraphics[height = 3.8cm, width = 5cm]{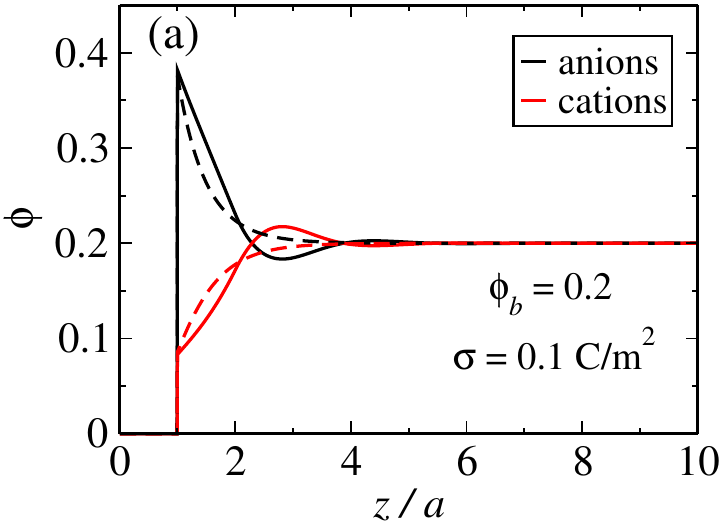}
    \includegraphics[height = 3.8cm, width = 5cm]{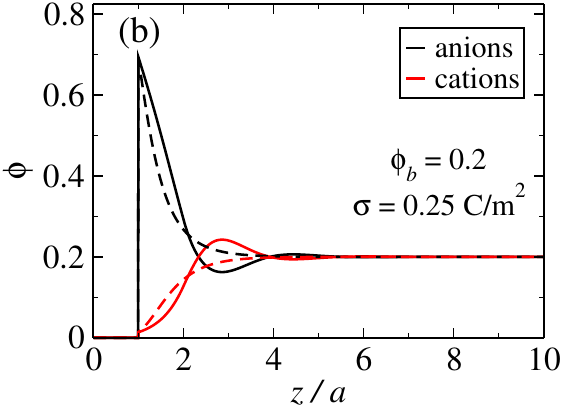}\\
    \includegraphics[height = 3.8cm, width = 5cm]{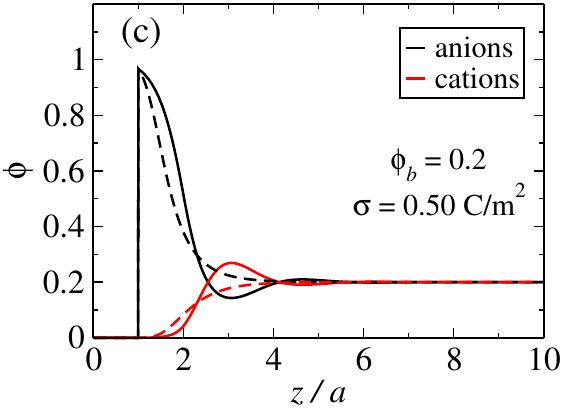}
    \includegraphics[height = 3.8cm, width = 5cm]{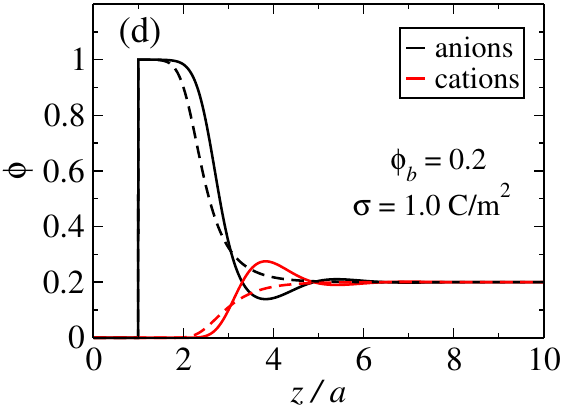}\\ 
     \includegraphics[height = 3.8cm, width = 5cm]{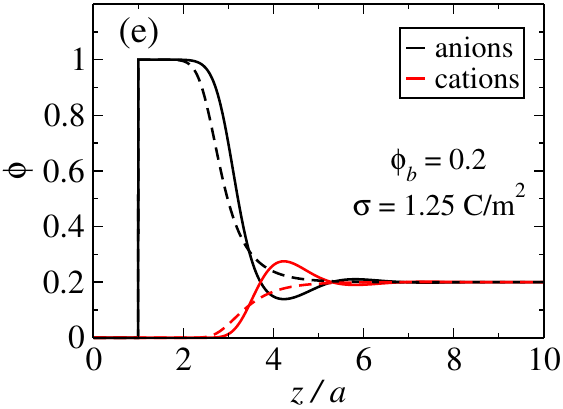}
     \includegraphics[height = 3.8cm, width = 5cm]{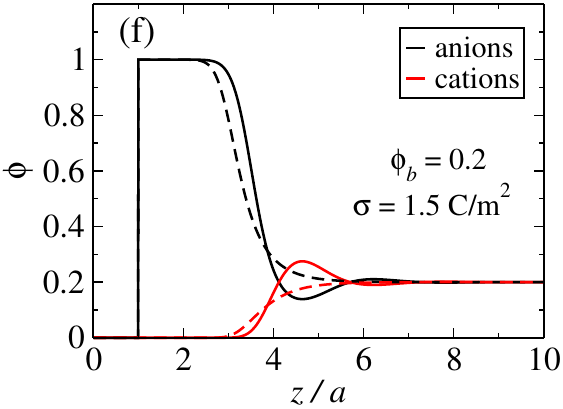} 
    \caption{Local packing fractions of anions (black curves) and cations (red curves) around a flat catode of different surface charges, for an electrolyte with dielectric constant $\varepsilon_r=80$. Solid lines are results from the mPF approach, whereas dashed lines are predictions from the standard PF model. In all cases, the bulk packing fraction is fixed at $\phi_b=0.2$, and the surface charge densities are displayed in the plots.}
    \label{fig:fig3}
\end{figure}

The structure properties of the EDL in the case of an aqueous electrolyte, $\lambda_B=0.72$ nm, in the vicinity of a cathode located at $z=0$ at various surface charges, are summarized  in Fig.~\ref{fig:fig3}. Solid lines are results from the mPF approach, whereas dashed lines stand from standard PF predictions. A clear distinction in qualitative behaviours is observed in all cases between the two approaches. While the PF profiles for counterions (coions) decay (increase) monotonically to their bulk values, there is a clear minimum in counterion distribution (together with a peak in coion profiles) in the case of mPF, before relaxing to the bulk concentrations. In all cases, the layer of counterions around the electrode is followed by a second layer of accompanying coions. Such a layering effect can be clearly attributed to the presence of fluctuations in the averaged potential due to the exclusion cavity around ions at the EDL. Note that in the new approach the counterion saturation effect near the charged wall is not only present, but is rather enhanced with respect to the standard PF approach. At low-to-moderate surface charges, the counterions display a sharp decay close to the interface. As the charge increases, the local packing fraction approaches unity (see Fig. \ref{fig:fig3}c), and the counterions begin to progressively fill in all sites at the layers closest to the wall.  The slower decay in counterion distribution near wall in the mPF results from the loss in the counterion ability to locally screen the surface charge, as placing a counterion at the ELD also requires the creation of a cavity that reduces the potential at the ion center. Such an increase in the amount of counterions ``lining up'' at the interface indicates the presence of  an excess charge at the first layer that exceeds the surface charge, a phenomenon largely known in physical-chemistry community as \textit{overcreening}. As a consequence, an adjacent layer of opposite charge is developed (corresponding to the local maxima in coion concentration), until the local electroneutrality is achieved. 

In order the further analyse the interplay between overscreening and charge ordering, it is convenient to define an integrated charge per unit of transversal area $A$ as
\begin{equation}
\sigma_{liq}(z)=\sigma+\dfrac{1}{v}\int_0^{z}\left(\phi_{+}(z)-\phi_{-}(z)\right)dz.
\label{sigma_z}
\end{equation}

\begin{figure}[h!]
    \centering
    \includegraphics[height = 3.8cm, width = 5cm]{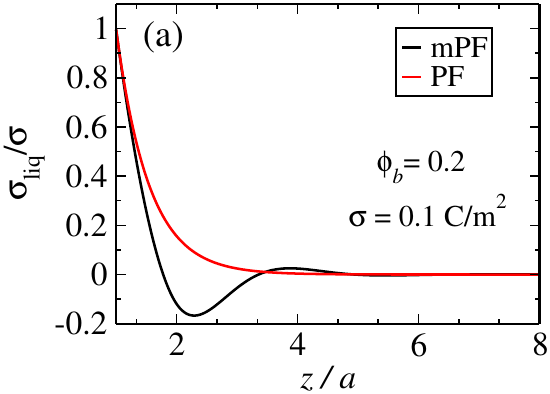}
    \includegraphics[height = 3.8cm, width = 5cm]{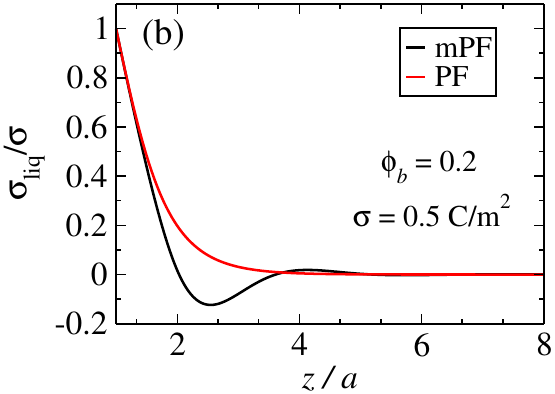}\\
    \includegraphics[height = 3.8cm, width = 5cm]{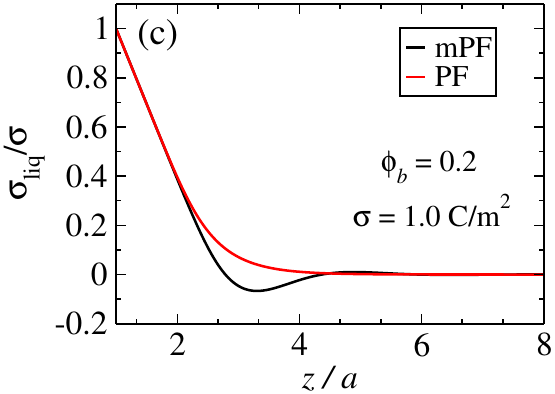}
    \includegraphics[height = 3.8cm, width = 5cm]{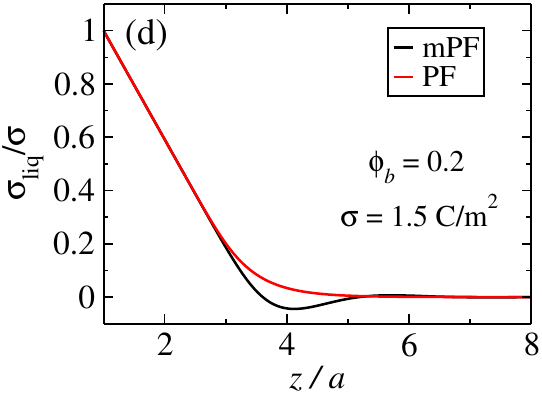}
    \caption{Integrated surface charges at a distance $z$ from the the positively charged electrode, computed from Eq. (\ref{sigma_z}) at different surface charges and a fixed packing fraction of $\phi_b=0.2$. Black lines are results from the mPF, while the red lines are predictions from standard PF model. }
    \label{fig:fig4}
\end{figure}

This quantity measures the net surface charge accumulated at a distance $z$ from the charged catode (including its own surface charge), and is therefore appropriate to relate charge ordering and screening at the EDL. Figure \ref{fig:fig4} shows the net integrated charge per unit area for the aqueous electrolyte at different surface charges, resulting from the mPF (black lines) and PF (red lines) models. While the standard PF model predicts a monotonic screening of the charge surface by the intervening counterions, the net charge in the mPF model clearly features an overscreening. The amount of condensed counterions has a net charge that over-compensates the surface charge, leading to charge reversal. This mechanism is clearly more pronounced in the case of low surface charges, where the net charge displays a second changing in sign due to charge layering (see Fig. \ref{fig:fig3}). The degree of overscreening is clearly reduced with increasing of the surface charge, as counterion saturation (overcrowding) starts to become the dominant effect at the EDL. In this regime, a large layer of counterions has to be packed at the wall in order to neutralize its charge, thus precluding the emergence of charge ordering. As a consequence, the thickness of the double layer becomes larger as $\sigma$ increases. This mechanism is responsible for the decrease in differential capacitance at larger applied voltages, in contrast to the unbound increasing observed in PB-based approaches, where the double layer becomes increasingly thinner at larger surface charges.

\begin{figure}[h!]
    \centering
    \includegraphics[height = 3.8cm, width = 5cm]{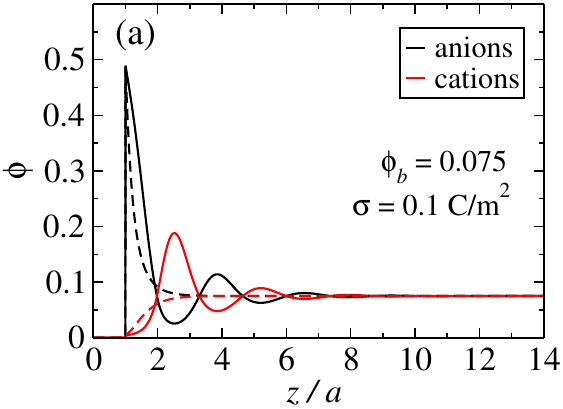}
    \includegraphics[height = 3.8cm, width = 5cm]{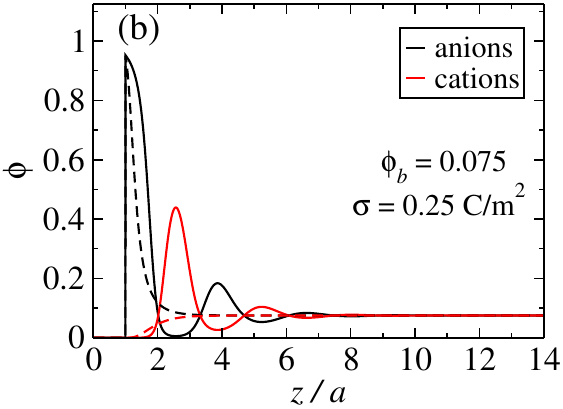}\\
    \includegraphics[height = 3.8cm, width = 5cm]{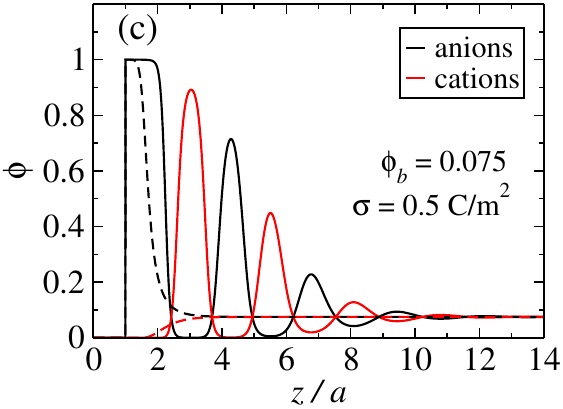}
    \includegraphics[height = 3.8cm, width = 5cm]{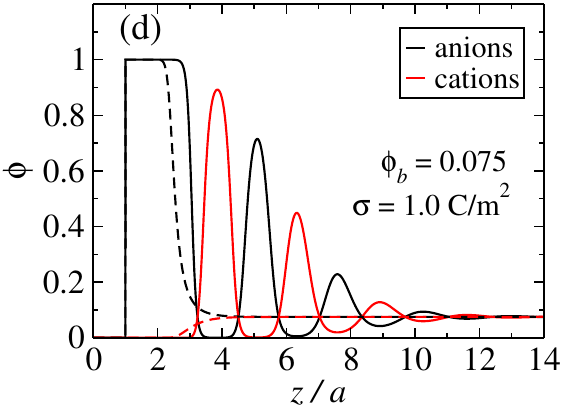}\\
     \includegraphics[height = 3.8cm, width = 5cm]{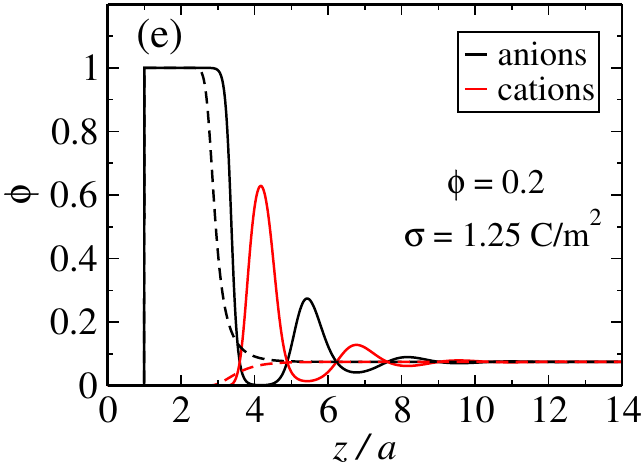}
     \includegraphics[height = 3.8cm, width = 5cm]{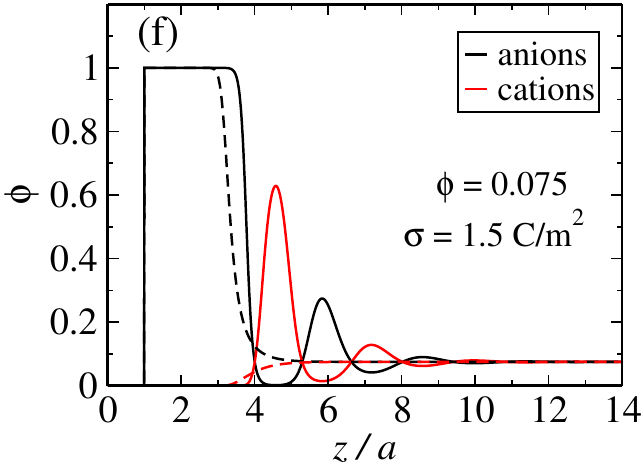} 
    \caption{Local packing fractions of anions (black curves) and cations (red curves) around a flat catode of different surface charges, for an electrolyte with dielectric constant $\varepsilon_r=18$. Solid lines are results from the mPF approach, whereas dashed lines are predictions from the standard PF model. In all cases, the bulk packing fraction is fixed at $\phi_b=0.075$, and the charge densities are displayed in the plots.}
    \label{fig:fig5}
\end{figure}

In spite of the qualitative differences between the structure of the EDL observed between PF and mPF models, the weak couple regime typical of aqueous 1:1 electrolyte does not allow for the emergence of strong charge ordering. In order to investigate the ability of the proposed mPF in predicting charge layering effects, we now proceed to investigate the structure of the EDL for the case of a much smaller dielectric constant $\varepsilon_r=18$, most typical of solvent-free electrolytes. This is done in Fig.~\ref{fig:fig5}, which compares ionic profiles from both mPF and PF models at different surface charges, for a fixed bulk packing fraction of $\phi_b=0.075$. As before, the counterion distributions in the PF model decays monotonically, and counterion saturation once again takes place at strongly charged surfaces. This behavior differs drastically from the one observed in the mPF approach, specially in the regime of small/moderate surface charges, where the EDL structure is dominated by strong charge layering. Such charge oscillations amount to an inefficient screening of the surface charge by the underlying electrolyte, as charge local electroneutrality is only achieved many particle layers away from the interface. These predictions are in qualitative agreement with the structure of real ionic liquids in the vicinity of charged electrodes, reported both in experiments~\cite{Wei17,Zhang24} and computer simulations~\cite{Souza20,Kon23}. The emergence of such long range charge ordering is usually assigned to strong ionic correlations that overweight configurations in which oppositely charged particles are brought close together~\cite{Levin02}.   Taking proper care of such strong correlations (which are fully neglected at the mean-field level) is usually a difficult task. In this sense, it is quite remarkable that a simple modification in the mean-field PF approach -- namely the concomitant introduction of charged cavities alongside exclusion volume -- is able to qualitatively capture the layering effect up to large distances from the interface.

As in the case of simple electrolytes, the charge layering effect in this case is also related to an ``overpacking'' of counterions at the first layers after the interface. The net surface charge is thus reversed, leading to the formation of subsequent charge layers that attempt to neutralize the previous ones. The number of such layers and their relative amplitudes depends non trivially on the cathode charge. All these trends are summarized in Fig.~\ref{fig:fig6}, in which the net integrated charge at a distance $z$ from the cathode is displayed for different surface charges. The degree of overscreening at the adjacent layer is more pronounced in the case of weakly charged surfaces, where the cathode charge is reversed to nearly half of its value. Nevertheless, the range of the screening effect is roughly independent on the cathode charge, as the first layer of counterions becomes deeper at larger charges in virtue of counterion saturation. Again, this is in strong contrast with results obtained from PF model. The large deviations between the two models can be attributed to the weaker ability of a counterion to screen an external charge when a cavity is placed around its center.

\begin{figure}[h!]
    \centering
    \includegraphics[height = 3.8cm, width = 5cm]{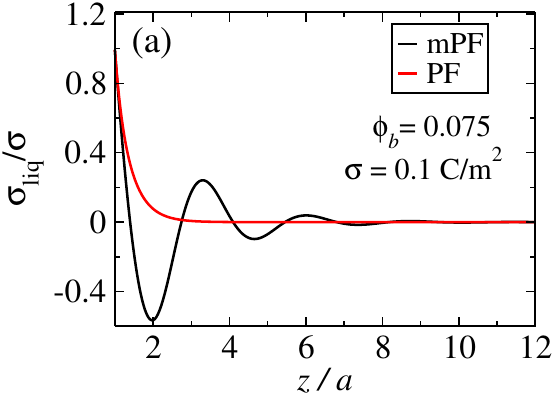}
    \includegraphics[height = 3.8cm, width = 5cm]{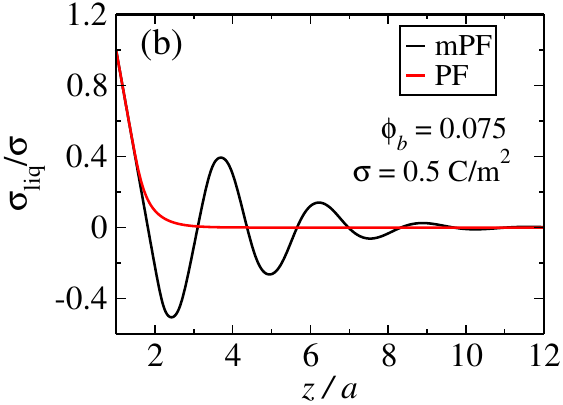}\\
    \includegraphics[height = 3.8cm, width = 5cm]{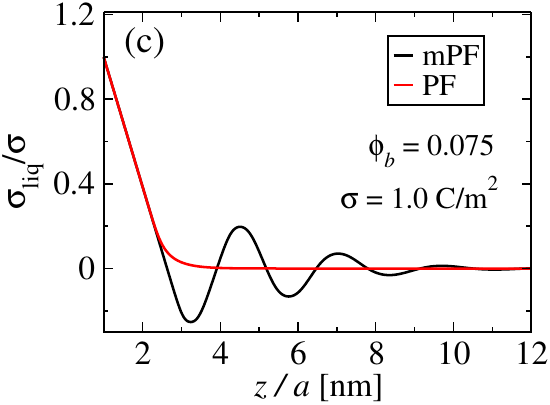}
    \includegraphics[height = 3.8cm, width = 5cm]{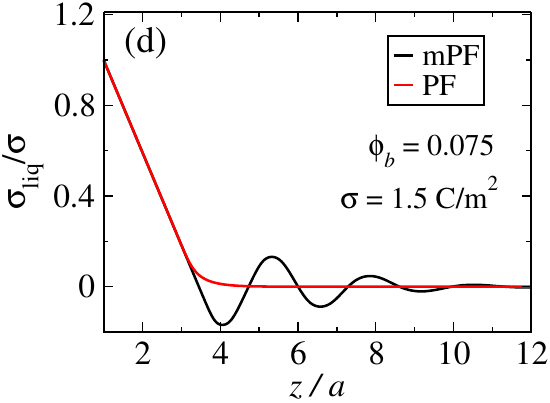}
    \caption{Integrated surface charges at a distance $z$ from the the positively charged electrode, computed from Eq. (\ref{sigma_z}) at different surface charges and a fixed packing fraction of $\phi=0.075$, for an electrolyte with dielectric constant $\varepsilon=18$. Black lines are results from the mPF, while the red lines are predictions from standard PF model. }
    \label{fig:fig6}
\end{figure}

\subsection{Differential capacitance}

It is clear from the previous analysis that PF and mPF models predict quite distinct structures for the EDL, particularly in the strong coupling limit of low permittivity electrolytes. It is not clear yet whether such qualitative differences in EDL structure will play a significant role in shaping capacitance curves at different regimes. These curves provide a convenient measure of the EDL ability to store electrostatic energy at varying applied voltages. On the other hand, the overall behavior of capacitance curves is known to be strongly model-dependent, as the differential capacitance is influenced by the fine details of EDL structure in a non-trivial way. Despite these difficulties, a simple estimation of the dependence of the  differential capacitance on the EDL thickness $L$ (defined as the distance beyond which the charges on the plate become fully screened) can be done as follows.   The differential capacitance is defined as $C_{diff}=\dfrac{d\sigma}{dV}$, where $V=\psi(0)-\psi_b$ is the potential drop between the electrode and the bulk solution (where the potential is $\psi_b$). Since the electric field in the bulk $E_b$ is zero, this quantity should scale as $V\sim L\bar{E}$, where $\bar{E}$ is the averaged electrostatic field. If the field decays monotonically away from the surface (as is the case in the PF approach), a rough estimation for $\bar{E}$ shows that it is proportional to the field on the plate, $\bar{E}\sim \sigma/\varepsilon$, in such a way that $C_{diff}$ should scale as $C_{diff}\sim \sigma/V \sim \varepsilon/L$. Decreasing the double layer thickness thus leads to an increase in differential capacitance. This explains the typical ``U''-shape behavior of the capacitance curve obtained from the PB approach~\cite{Fedo14}, as the absence of size correlations leads to an increasingly thinner counterion layer as $\sigma$ increases. Such a unbounded growth in capacitance at large voltages is properly corrected within the PF framework, as counterion saturation leads to an increase in EDL thickness $L$ at large $\sigma$. We note, however, that such a rough estimation does not apply directly to the mPF approach, as the charge layering shown in Figs.~\ref{fig:fig4} and \ref{fig:fig6} implies that the electric field changes sign over different cycles across the ELD, such that no clear scaling law can be devised between $\bar{E}$ and $L$.

\begin{figure}[h!]
    \centering
    \includegraphics[height = 4cm, width = 5.8cm]{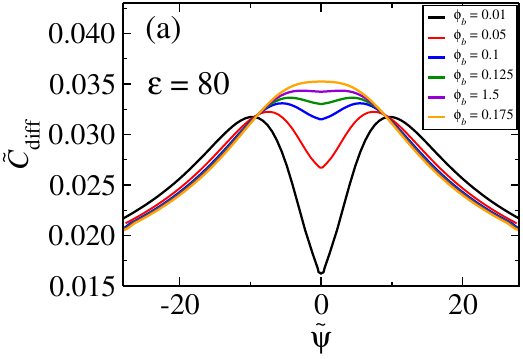}
    \includegraphics[height = 4cm, width = 5.8cm]{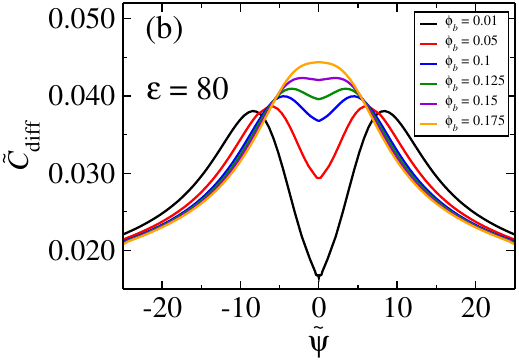}\\
    \includegraphics[height = 4cm, width = 5.8cm]{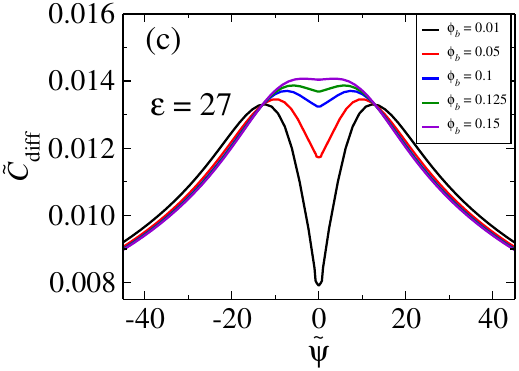}
    \includegraphics[height = 4cm, width = 5.8cm]{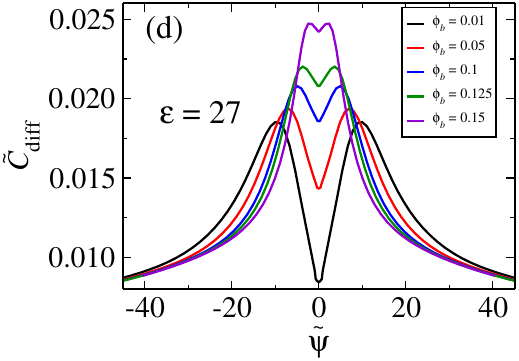}\\
    \includegraphics[height = 4cm, width = 5.8cm]{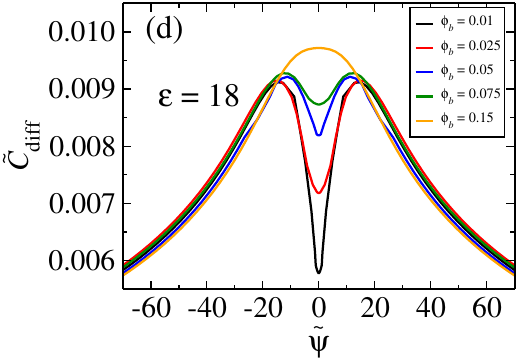}
    \includegraphics[height = 4cm, width = 5.8cm]{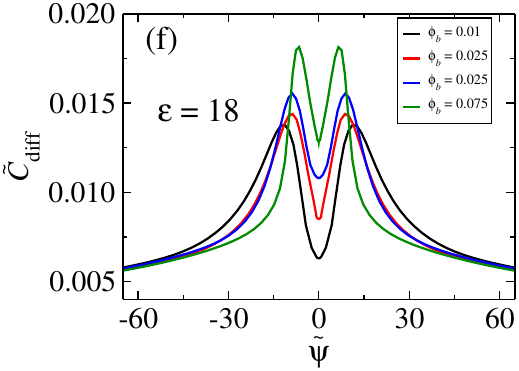}
    \caption{Dimensionless differential capacitance $\tilde{C}_{diff}\equiv \lambda_BC_{diff}/\varepsilon$ as a function of the dimensionless wall potential $\tilde{\psi}=\beta q \psi(0)$ at several bulk packing fraction of cations, $\phi_b$. Left panels are results from the traditional PF model, while results from the mPF approach are presented in the right panels. The electrolyte dielectric constants are $\epsilon=80$ (top panels), $\varepsilon=27$ (middle panels) and $\varepsilon=18$ (bottom panels).  }
    \label{fig:fig6}
\end{figure}

Comparison between capacitance curves calculated from both PF and mPF models is shown in Fig.~\ref{fig:fig6} for three representative polarities: $\varepsilon_r=80$ (upper panels), $\varepsilon_r=27$ (middle panels) and $\varepsilon_r=18$ (lower panels). Left panels stand for PF results, whereas predictions from mPF are displayed in the right panels. A similar trend is observed in all cases, with the capacitances displaying camel-like shapes at small packing fractions.  As $\phi_b$ increases, the peaks start to come close together, eventually merging to provide a bell-like shape at sufficiently high $\phi_b$. In spite of the discrepancies observed regarding the EDL structure, the overall qualitative behavior of capacitance curves is remarkably similar. In particular, the transition from camel to bell-like shape seems to take place close to $\phi_b\approx 0.15$ for large ($\varepsilon_r=80$) and intermediate ($\varepsilon_r=27$) polarities, in both models. We notice, however, that the values of $C_{diff}$ are larger than those predicted by the PF approach. Moreover, the camel ``humps'' are more pronounced in the mPF, and lie closer to one another. This is likely to be due to a different nature of the growth of $C_{diff}$ at small surface potentials. While in the PF approach the increase of $C_{diff}$ can be assigned to a decrease in EDL thickness at small $\sigma$ (where counterion saturation has not yet set in), the width of the EDL in the mPF model actually increases at small potentials due to the onset of charge layering. In this case, the sharp increase in differential capacitance at small $\psi$ is most likely to be attributed to an increase in the averaged field $\bar{E}$ in virtue of charge oscillations. In the strong coupling regime of small $\varepsilon$, the strong charge oscillations at large packing fractions extends all the way into the bulk electrolyte, rendering the numerical solutions at fixed bulk concentrations unstable. As we shall shortly see, such numerical instability is due to a unbound increase in the dumping range of charge oscillations in these regimes. This leads to very long range oscillations, thus requiring numerical integration over an increasingly large region beyond the wall in order to suppress charge layering. Unfortunately, numerical calculations for capacitances within the mPF become unstable beyond $\phi_b\approx 0.08$ in the case where $\varepsilon_r=18$ (see Fig.~\ref{fig:fig6}f). However, comparison between the overall shapes of $C_{diff}$ in Figs.~\ref{fig:fig6}b,  \ref{fig:fig6}c and \ref{fig:fig6}c clearly suggests a tendency of shifting the camel-bell transition towards larger packing fractions in the mPF model at strong couplings. 

\subsection{Charge layering instability} 

\begin{figure}[h!]
    \centering
    \includegraphics[height = 4.5cm, width = 6cm]{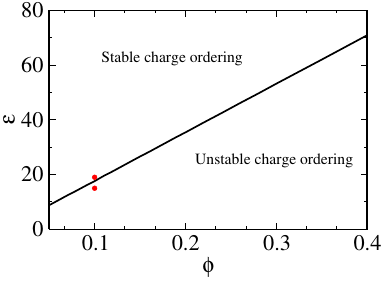}
    \caption{Transition line that delimits the region of charge ordering stability in the $(\phi_b,\varepsilon)$ plane, considering a correlation cavity whose size equals the ionic size, $d=r_{ion}$. In the unstable region, the decay length of charge oscillations diverges, and charge layering is extended over the entire system. }
    \label{fig:fig7}
\end{figure}

\begin{figure}[h!]
    \centering
    \includegraphics[height = 3.8cm, width = 5.5cm]{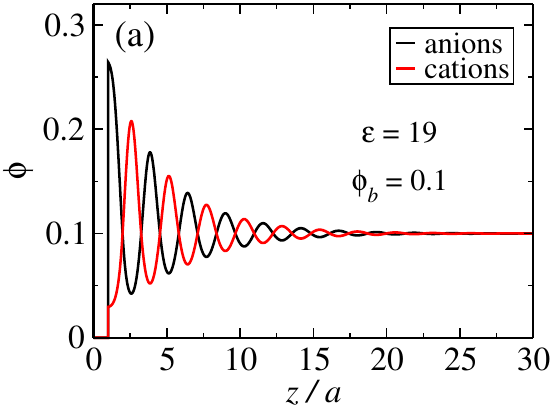}
    \includegraphics[height = 3.8cm, width = 5.5cm]{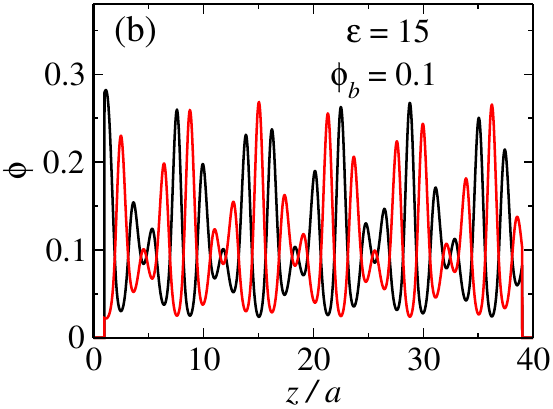}
    \caption{Transition line that delimits the region of charge ordering stability in the $(\phi_b,\varepsilon)$ plane, considering a correlation cavity whose size equals the ionic size, $d=r_{ion}$ and a fixed surface charge of $\sigma=0.05$ C/m$^2$. In the unstable region, the decay length of charge oscillations diverges, and charge layering is extended over the entire system. }
    \label{fig:fig8}
\end{figure}

We now address the question of charge layering stability in the mPF model, with the aim of delimiting the range of validity of the model over the parameter space. As shown in Section~\ref{linear}, the effective screening length -- conceived as the inverse dumping parameter that controls the decay of charge oscillations in the linear regime -- becomes larger as the roots $\tilde{k}_0$ of Eq. (\ref{roots}) in the complex plane approach the real axis. Charge oscillations then become long-ranged, eventually reaching a regime of undumped oscillations when the roots $\tilde{k}_0$ are real. Such a situation can be physically achievable only in cases of small system sizes, such as electrolytes in porous media or confined between electrodes. For open systems, a bulk phase of vanishing electric field must be established far away from external sources, such that the undumped oscillations loose any physical significance. It is therefore important to clearly distinguish the regions in parameter space where such instabilities might take place, thereby  delimiting the range of validity of the proposed approach in case of non-confined systems. Since the Debye length $\tilde{\kappa}=\sqrt{8\pi\lambda_B\phi_b d^2/v}$ in Eq.~(\ref{roots}) depends on the bulk packing fraction $\phi_b$ and the degree of polarity $\varepsilon$, it is possible to identify the set of points $(\phi_b,\varepsilon)$ for which  Eq. (\ref{roots}) admits real-valued solutions. This is done in Fig.~\ref{fig:fig7}, which shows the transition line separating stable (above the line) and unstable (below the line) regimes, considering a fixed correlation cavity of size $d=r_{ion}$. At small packing fractions, the stability zone is extended over the entire set of polarizabilities. At high concentrations, the system becomes unstable at permitivitties typical of a strong coupling regime. The crossover between unstable and stable charge ordering is clearly illustrated in Fig.~\ref{fig:fig8}, in which density profiles are shown for two representative systems as the transition line is crossed (corresponding to the points marked in red in Fig.~\ref{fig:fig7}). Despite the small difference in dielectric constants (from $\varepsilon_r=19$ to $\varepsilon_r=15$), a drastic difference in qualitative behavior is observed in both cases. The envelope of charge oscillations is changed from a dumped decay to one featuring a harmonic shape typical of an interference pattern. In the case of a confined electrolyte with fixed concentration, such a situation implies that the overall shape and number of oscillating cycles should depend on the system size. Moreover, it renders the capacitance curves of such systems ill-defined, due to abrupt changes in ionic profiles (as new oscillating modes start to emerge by increasing the surface charges). Such singularities in the capacitance behavior of ionic liquid electrolytes have been reported in the literature~\cite{Liu06}, and have been indeed attributed to the onset of strong charge ordering~\cite{Limm15}. We notice, however, that the emergence of such behaviour in the proposed mPF depends crucially on the shape and size attributed to the electrostatic cavities. In most realistic approximations, the cavities should be taken as smooth functions whose decay depends on the local charge environment.

\section{Conclusions}\label{close}

A model has been proposed that improves the tradition PF approach \textit{via} the addition of correlation cavities that mimics the absence of charge overlapping in a test-particle insertion. In spite of its simplicity, this modification leads to dramatic differences in the description of EDL structure. In strong coupling regimes, the proposed mPF approach predicts an enhanced charge ordering effect, which is absent in the original PF formulation. The physical mechanism responsible for this effect is the so-called overscreening, in which the amount of counterions packed closed to the electrode is large enough to effectively reverse its charge. As a result, further layers of different charges are developed until full screening is achieved. On the other hand, the overscreening can be attributed to a reduced ability of the electrolyte to screen an external charge due to the charge cavity surrounding a tagged ion. At large surface charges, overcrowding effects start to dominate over charge ordering, leading to counterion saturation similar to the one observed in the PF approach. Despite the strong deviations regarding the EDL properties, both PF and mPF models predict similar qualitative behaviors for the capacitance curves at different couplings and ionic strengths. In particular, the transition from camel shaped to bell-like curves occurs at similar packing fractions. At sufficiently strong couplings, the proposed mPF theory becomes unable to describe electrolytes with well-defined bulk phases. This is due to the emergency of a unbounded charge ordering, which is extended over the whole system. Using a linear response theory, we could predict the range of parameters in which such unstable solutions start to take place. We notice, however, that such unscreened charge layering can represent a realistic effect in cases of confined electrolytes. 

The proposed approach has room for improvements in various directions to describe different aspects inherent to different types of electrolytes and RTIL. For instance, charge asymmetry between cations and anions is a key ingredient of many RTIL, which can be easily incorporated into the model by considering cavities of different sizes. Besides, different sizes can be assigned to charge and size cavities. While the size cavities should scale with the particle size, the charge cavity is more closely related to particle correlations. It can be taken, for example, to scale with the averaged particle distance, in which case it should depend non-trivially on the local concentrations. In a more realistic situation, the charge cavity should depend on the pair of ions under consideration, as the averaged distance between cations and anions should be much smaller than that of equally charged ions. All these aspects can be easily incorporated into the mPF model. Due to its physical transparency and ease of implementation, we expect that the proposed model can be further extended to shed light on different mechanisms governing the behaviour of RTIL close to electrified interfaces.  





%

\end{document}